\documentclass[12pt] {iopart}
\usepackage[dvips]{graphicx}

\newcommand{\be}{\begin{equation}}
\newcommand{\ee}{\end{equation}}
\newcommand{\ba}{\begin{eqnarray}}
\newcommand{\ea}{\end{eqnarray}}

\newcommand{\mb}{\mathbf}

\begin{document}

\title{On Non-Abelian Holonomies }
\author{J. Alfaro \dag \footnote[3]{jalfaro@puc.cl}
,H.A. Morales-T\'ecotl\ddag,M. Reyes \dag \ and  L.F. Urrutia\P}

\address{\dag Facultad de F\'\i sica\\
Pontifica Universidad Cat\'olica de Chile\\
Casilla 306, Santiago 22, Chile }
\address{\ddag Departamento de F\'\i sica\\
Universidad  Aut\'onoma Metropolitana-Iztapalapa\\
A. Postal 55-534, M\'exico D.F. 09340, M\'exico\\
}
\address{\P Departamento de F\'\i sica de Altas Energ\'\i as\\
Instituto de Ciencias Nucleares\\
Universidad Nacional Aut\'onoma de M\'exico\\
A. Postal 70-543, M\'exico D.F. 04510, M\'exico \\}

\begin{abstract}
We provide a method and the results  for the calculation of  the holonomy of a Yang-Mills connection in
an  arbitrary triangular path, in an expansion (developed here to fifth order) in powers of the corresponding
segments. The results might have applications in generalizing to Yang-Mills fields previous calculations  of
the corrections to particle dynamics induced by loop quantum gravity,  as well as in the field of random lattices.
\end{abstract}


\maketitle


\section{Introduction}

Constraints in  Lorentz covariance violations have been experimentally studied since a
long time ago
\cite{HUDR} by obtaining observational bounds upon the violating parameters. Recent experiments have shown an impressive increase in their sensitivities  thus producing even more stringent bounds \cite{Exper2}.

In order to correlate such experimental results, Kostelecky and collaborators
have proposed a phenomenological extension of the standard model,
which incorporates the most general Lorentz-violating interactions compatible with power counting renormalizability together with the particle structure of the standard model \cite{kostelecky1}. An impressive  number of applications to very different processes have been already considered,
as can be seen in Ref. \cite{CPTPROC}, for example.

Different models accounting for minute Lorentz violations have recently arisen in
the context
of quantum gravity induced corrections to the propagation and interactions of
particles \cite{Gambini,amufot,Us,Ellis,QuantFluct}. This amounts to  realize the generic violating
parameters appearing in the standard model
extension in terms of specific quantities involving the Planck length $\ell_P$ together with additional  physically relevant objects. Moreover, the high precision  obtained in the
determination of the experimental bounds has brought  quantum gravity
induced effects to the level of observable phenomena \cite{BILLER,SUV,ap,observ}.

On the other hand, in the early 80's, the growing importance of computer simulations of gauge theories
required a short distance cutoff of geometrical origin such as a lattice.
However, regular lattices
break essential symmetries of continuum theories such as translational and rotational invariance.
Motivated by the need of maintaining these symmetries, the field theory on a random lattice was suggested
\cite{cfl}. Further on, the connection of random lattices with quantum gravity and strings
was studied and low dimensional systems on random lattices were solved using
matrix model techniques \cite{jan}.

In this work we concentrate on some aspects arising in the process of generalizing
the loop quantum gravity inspired model described
in \cite{amufot,Us}  to Yang-Mills fields, in order to obtain  the non-Abelian generalization of the
corrections previously found for the dynamics of
photons. Namely, corrections to standard matter dynamics are obtained by means of calculating non-Abelian holonomies, either of gravitational or Yang-Mills type, around triangular paths. To this end
we have to revise  and extend the procedure of Ref.\cite{FGK} that was applied to the case of rectangular cells.
It turns out that the method we present in this paper contains as a particular case
the result of Ref.\cite{FGK}, though it is applicable to arbitrary cells made up of triangles.
The basic building block in our analysis is the holonomy along a straight line
segment, which  characteristic property of been path-ordered is   consistently maintained to all orders in our expansion.

The problem we deal with here is closely related to the non Abelian Stokes theorem which has been repeatedly
discussed in the literature \cite{FGK,STOKES}.

This paper is organized as follows: in Section 2 we state the problem to be dealt with and introduce some
notation. Section 3 summarizes  the results for the Abelian case which we intend to generalize here.
The non-Abelian case is  subsequently discussed in Section 4 which contains our main results.
Using the procedure  of Ref. \cite{FGK} the corresponding calculation  is performed in Section 5,
which allows us to show some discrepancies  that arise  between the two methods. Finally we close with
a summary and discussion in Section 6.

The computations  involved in deriving ${\mb{h}}_{\alpha_{IJ}}^{(5)} $ were done using FORM\cite{form}.

\section{Statement of the problem}

\

The proposed method to obtain the quantum gravity induced corrections to the Yang-Mills Lagrangian
requires the calculation of the object
\begin{equation}
\label{TGHOL}
T_\rho=tr\left({\mathbf G}_\rho\,{\mathbf h}_{\alpha_{IJ}} \right),
\end{equation}
where ${\mathbf G}_\rho$ are the generators of the corresponding Lie algebra and ${\mathbf h}_{\alpha_{IJ}}$ 
 is the holonomy of the Yang-Mills connection ${\mathbf  A}_a=A_a^\rho\,{\mathbf G}_\rho$ in the triangle $\alpha_{IJ}$, with vertex  $v$, defined
by the vectors ${\vec s}_I$ and ${\vec s}_J$, arising from the vertex $v$ in the way described below.

 Our  main task will be to construct an expansion of $T_\rho$ in powers of the segments $s^a_I, \,
s^b_J$.

To be more precise, we have
\be
{\mathbf h}_{\alpha_{IJ}}=P\,exp\left({\oint_{\alpha_{IJ}} {\mathbf A}_a({\vec x}(s))\,\frac{dx^a}{ds}\,ds}\right),
\ee
where $P$ is a path-ordered product to be specified later.
As shown in Fig.1, the closed path $\alpha_{IJ}$, parameterized by ${\vec x}(s)$, is defined in the following  way: we start from the vertex $v$ following a straight line with the direction and length of ${\vec s}_I$, then follow another straight line in the direction and length of ${\vec s}_J-{\vec s}_I$, and finally return to $v$ following $-{\vec s}_J$. From the definition of the holonomy, we have the transformation property
\be
{\mathbf h}_{\alpha_{IJ}}\rightarrow  {\mathbf U}(v)\,{\mathbf h}_{\alpha_{IJ}}\,{\mathbf U}(v)^{-1},
\ee
under a gauge transformation of the connection, where ${\mathbf U}(v)$ is a group element evaluated at the vertex $v$. In other words, ${\mathbf h}_{\alpha_{IJ}}$  transforms covariantly under the group.

\section{The Abelian case}

The corresponding calculation was performed in Ref.\cite{amufot} and
here we summarize the results in order to have the
correct expressions to which the non-Abelian result must reduce when taking the commuting limit. In this case
Eq.(\ref{TGHOL}) reduces to
\be
T=exp(\Phi_{IJ})-1 ,
\ee
where $\Phi_{IJ}$ is the magnetic flux through the area of the triangle, given by
\begin{eqnarray}
\Phi ^{B}(F_{IJ})&=&\oint_{\alpha _{IJ}}dt\;\dot{s}^{a}(t)A_{a}(t) \nonumber \\
&=&\int_{{\vec{v}}}^{{\vec{v}}+{\vec{s}}_{I}}{{A}}%
_{a}\,dx^{a}+\,\int_{{\vec{v}}+{\vec{s}}_{I}}^{{\vec{v}}+{\vec{s}}_{J}}{%
{A}}_{a}\,dx^{a}+\int_{{\vec{v}}+{\vec{s}}_{J}}^{{\vec{v}}}{%
{A}}_{a}\,dx^{a}, \label{STOKES}
\end{eqnarray}
where the connection $A_a({\vec x}(s))$ is now a commuting object.

The basic building block in (\ref{STOKES}) is
\begin{eqnarray}
\int_{{\vec{v}}_{1}}^{{\vec{v}}_{2}}{{A}}_{a}({\vec{x}})\,dx^{a}
&=&\int_{0}^{1}{{A}}_{a}({\vec{v}}_{1}+t\,({\vec{v}}_{2}-{\vec{v}}%
_{1}))\,({\vec{v}}_{2}-{\vec{v}}_{1})^{a}\,dt  \nonumber \\
&=&\int_{0}^{1}{{A}}_{a}({\vec{v}}_{1}+t\,\vec{\Delta})\,\Delta
^{a}\,dt  \nonumber \\
&=&\left( 1+\frac{1}{2!}\Delta ^{b}\partial _{b}+\frac{1}{3!}(\Delta ^{b}\partial
_{b})^{2}+\dots \right) \Delta ^{a}{A}_{a}(v),
\end{eqnarray}
with$\;\Delta ^{a}=({\vec{v}}_{2}-%
{\vec{v}}_{1})^{a}$. The infinite series in parenthesis is
\begin{equation}
F(x)=1+\frac{1}{2!}x+\frac{1}{3!}x^{2}+\frac{1}{4!}x^{3}+\dots =\frac{e^{x}-1%
}{x},  \label{EFE}
\end{equation}%
yielding
\be
\int_{{\vec{v}}_{1}}^{{\vec{v}}_{2}}{{A}}_{a}({\vec{x}}%
)\,dx^{a}=F(\Delta ^{a}\,\partial _{a})\left( \Delta ^{a}\,{{A}}%
_{a}({\vec{v}}_{1})\right) .\quad
\label{PATHI}
 \ee
In the following we employ the notation $\Delta
^{a}\,V_{a}={\vec{\Delta}}\cdot {\vec{V}}$. Using the above result in the three
integrals appearing in (\ref{STOKES}) and after some algebra, we obtain
\begin{eqnarray}
\Phi ^{B}(F_{IJ}) &=&F_{1}({\vec{s}}_{I}\cdot {\nabla},{\vec{s}}%
_{J}\cdot {\nabla})\,\,s_{J}^{a}\,s_{I}^{b}\,\left( \partial _{a}\,{%
{A}}_{b}({\vec{v}})-\partial _{b}\,{{A}}_{a}({\vec{v}}%
)\right) \nonumber \\
&=&F_{1}({\vec{s}}_{I}\cdot {\nabla},{\vec{s}}_{J}\cdot {\nabla}%
)\,\,s_{J}^{a}\,s_{I}^{b}\,\epsilon _{abc}B^{c}(v),
\end{eqnarray}%
where the gradient acts upon the coordinates of ${\vec{v}}$. The function $%
F_{1}$ is \be
F_{1}(x,y)=\frac{y({\rm e}^{x}-1)-x({\rm e}^{y}-1)}{x\,y\,(y-x)}%
=-\sum_{n=1}^{\infty }\frac{1}{(n+1)!}\,\frac{x^{n}-y^{n}}{x-y}. \ee
Let us emphazise
that $F_{1}(x,y)$ is just a power series in the variables $x$ and $y$. Expanding in powers of the
 segments $s_{I}^{a}$ we obtain
\begin{eqnarray}
\Phi ^{B}(F_{IJ}) &=&\left( 1\,\,+\frac{1}{3}(s_{I}^{c}+s_{J}^{c})\,\partial
_{c}\,+\frac{1}{12}(s_{I}^{c}\,s_{I}^{d}+s_{I}^{c}\,s_{J}^{d}+s_{J}^{c}%
\,s_{J}^{d})\,\,\partial _{c}\,\partial _{d}+...\right) \times  \nonumber
 \\
&&\times \,\frac{1}{2}s_{I}^{a}s_{J}^{b}\epsilon _{abc}B^{c}(v).
\label{FLUX}
\end{eqnarray}%
Notice that the combination%
\be \frac{1}{2}s_{I}^{a}s_{J}^{b}\epsilon _{abc}={\cal A}\;n_{c},
\ee
is just the oriented area of the triangle with vertex $v$ and sides $%
s_{I}^{c},\;s_{J}^{c}$, joining at this vertex, having value ${\cal A}$ and unit
normal$\;$vector$\;n_{c}$.

To conclude we have to calculate

\ba
 \label{WLC}
\left(
e^{{{\Phi}}^{B}(F_{IJ})}-1\right) &=& \sum_{n=2}^{\infty}
\frac{1}{n!} ({{\Phi}}^{B}(F_{IJ}))^n= \sum_{n=2}^{\infty}
M_{nIJ},
\ea
where the subindex $n$ labels the corresponding power in the vectors $s^a$.
The results are
\ba \label{MJKD2}
M_{2IJ}:&=&  s^a_I s^b_J \frac{1}{2!}  {F}_{ab}, \\
\label{MJKD3}
M_{3IJ}:&=& s^a_I s^b_J \frac{1}{3!} (x_I+x_J)  {F}_{ab}, \\
\label{MJKD4}
M_{4IJ}:&=&s^a_I s^b_J  \frac{1}{4!} (x_I^2+x_Ix_J+x_J^2)  {F}_{ab}
+ s^a_I s^b_Js^c_Is^d_J \frac{1}{8}  {F}_{ab}                                 {F}_{cd},\\
\label{MIJD5}
M_{5IJ}:&=&\frac{1}{5!}(x_I^3+x_J^3+x_I^2x_J+x_Ix_J^2)s_I^as_J^b F_{ab}
\nonumber\\
&&+\frac{ s^a_I s^b_Js^c_I s^d_J}{4!}\left[(x_I+x_J){F}_{ab}{F}_{cd}+{F}_{ab}(x_I+x_J){F}_{cd}\right].
\ea
up to fifth order. We are using the notation $x_I={\vec
s}_I\cdot\nabla=s_I^a\,\partial_a$.

We expect that the non-Abelian generalization of the  quantities (\ref{MJKD2}), (\ref{MJKD3}), (\ref{MJKD4}), (\ref{MIJD5}) is produced by the replacement
\ba
&& A_a \rightarrow {\mathbf A}_a=A^\rho_a\,G_\rho, \quad \partial_a \rightarrow {\mathbf D}_a=
\partial_a-[{\mathbf A}_a , {\ } ], \\
&& F_{ab} \rightarrow {\mathbf F}_{ab}=
 \partial_a{\mathbf A}_b-\partial_b{\mathbf A}_a-[{\mathbf A}_a, {\mathbf A}_b].
\ea
Nevertheless, at this level there are potential ordering ambiguities which will be resolved in the next sections.

\section{The non-Abelian case}

In a similar way to the Abelian case we separate the calculation of the holonomy ${\mb{h}}_{\alpha_{IJ}}$ in three basic pieces through the straight lines along the sides of the triangle ${\alpha_{IJ}}$. We have
\be
{\mb{h}}_{\alpha_{IJ}}=P(e^{L_3})P(e^{L_2})P(e^{L_1})\equiv \,U_3\,U_2\,U_1,
\label{RHOL}
\ee
where
\ba
&&L_1=\int_0^ 1 dt\, {\mb{A}}_a({\vec v} + t \, {\vec s}_I)\, s^a_I,\\
&&L_2=\int_0^ 1 dt\, {\mb{A}}_a({\vec v}+ {\vec s}_I + t \, ({\vec s}_J-{\vec s}_I))\,(s^a_J-s^a_I),\\
&&L_3=\int_0^ 1 dt\, {\mb{A}}_a({\vec v}+ {\vec s}_J - t \, {\vec s}_J)\,(- s^a_J).
\ea

Here we have parameterized each segment with $0 \leq t \leq 1$.

\subsection{The basic building block}

Let us consider in detail the contribution
\be
U_1=P(e^{L_1}), \qquad L_1=\int_0^ 1 dt\, {\mb{A}}_a({\vec v} + t \, {\vec s_I})\, s_I^a,
\ee
with ${\vec s_I}= \{ s_I^a \}$.

Using the  definition
\begin{eqnarray}
U_1&=&1+\int_{0}^{1}dt {\mathbf A}_{a}({\vec v}+t{\vec s_I})s_I^{a}+\int_{0}^{1}dt\int_{0}^{t}dt'{\mathbf A}_{a}({\vec v}+t{\vec s_I}){\mathbf A}_{b}({\vec v}+t'\vec s_I)s_I^{a}s_I^{b}{}
\nonumber\\
&&{}+\int_{0}^{1}dt\int_{0}^{t}dt'\int_{0}^{t'}dt''{\mathbf A}_{a}{\mathbf A}_{b}{\mathbf A}_{c}s_I^{a}s_I^{b}s_I^{c}+ \dots,
\end{eqnarray}
for the path ordering, we arrive at the following expression
\begin{eqnarray}
\label{U1}
U_1&=&1+I_{1}(x){\mathbf A}_{a}(v)s_I^{a}+I_{2}(x,\bar{x}){\mathbf A}_{a}(v)\bar{{\mathbf A}}_{b}(v)s_I^{a}s_I^{b}\nonumber \\
&&+I_{3}(x,\bar{x},\bar{\bar{x}}){\mathbf A}_{a}(v)\bar{{\mathbf A}}_{b}(v){\bar{\bar{{\mathbf A}}}}_c(v)s_I^{a}s_I^{b}s_I^{c}+ \dots
\end{eqnarray}
Here we are adopting    the conventions
\begin{eqnarray}
&&x=s^{c}_{I}\partial_{c},\qquad \bar{x}=s^{c}_{I}\bar{\partial}_{c}, \qquad \bar{\bar{x}}=s^{c}_{I} \bar{\bar{\partial}}_{c},\\
&&I_{1}=F(x), \qquad
I_{2}=\frac{F(x+\bar{x})-F(x)}{\bar{x}},\\
&&I_{3}=\frac{1}{\bar{\bar{x}}}\left[\frac{1}{\bar{x}+\bar{\bar{x}}}( F(x+\bar{x}+\bar{\bar{x}})-F(x))-\frac{1}{\bar{x}}(F(x+\bar{x})-F(x))  \right].
\end{eqnarray}
with  $F(x)$ given by Eq.(\ref{EFE}).
The notation in Eq. (\ref{U1}) is that  each operator
 $x,\bar{x},\bar{\bar{x}}$ acts only in the corresponding field $A,\bar{A},\bar{\bar{A}}$ respectively.
We write
\be
U_1=\sum_N\, U_{1}^{(N)},
\label{U1N}
\ee
where the superindex  $N$ indicates the powers of $s_I^a$ contained in each term. A detailed calculation produces
\begin{equation}
U_{1}^{(1)}=s_{I}^{a}{\mb A}_{a}, \qquad U_{1}^{(2)}=\frac{1}{2}(x s_{I}^{a}{\mb A}_{a}+ s_{I}^{a}s_{I}^{b}{\mb A}_{a}{\mb A}_{b}),
\end{equation}
\begin{equation}
U_{1}^{(3)}=\frac{1}{3!}(x^{2} s_{I}^{a}{\mb A}_{a}+(\bar{x}+2x)s_{I}^{a}s_{I}^{b}{\mb A}_{a}\bar{{\mb A}_{b}}+s_{I}^{a}s_{I}^{b}s_{I}^{c}{\mb A}_{a}{\mb A}_{b}{\mb A}_{c}),
\end{equation}
\begin{eqnarray}
U_{1}^{(4)}&=&\frac{1}{4!}\left[x^{3}s_{I}^{a}{\mb A}_{a}+(3x^{2}+3x\bar{x}+\bar{x}^{2})s_{I}^{a}s_{I}^{b}{\mb A}_{a}\bar{{\mb A}_{b}}+(3x+2\bar{x}
+\bar{\bar{x}})s_{I}^{a}s_{I}^{b}s_{I}^{c}{\mb A}_{a}\bar{{\mb A}_{b}}\bar{\bar{{\mb A}_{c}}}\right.\nonumber \\
&&\qquad  \left.+ s_{I}^{a}s_{I}^{b}s_{I}^{c}s_{I}^{d}{\mb A}_{a}{\mb A}_{b}{\mb A}_{c}{\mb A}_{d}\right],
\end{eqnarray}
Specializing to the case ${\vec s}_I=(a,0,0)$ and to third order in $a$ we obtain
\be
\label{COMP}
U_{1}=1+a{\mb A}_{1}+\frac{1}{2}a^{2}(\partial_{1}{\mb A}_{1}+{\mb A}_{1}^{2})+\frac{1}{3!}a^{3}(\partial^{2}_{1}{\mb A}_{1}+{\mb A}_{1}\partial_{1}{\mb A}_{1}
+2(\partial_{1}{\mb A}_{1}){\mb A}_{1}+ {\mb A}_1^3).
\ee
Next we compare our result (\ref{COMP}) with the calculation according to the method of Ref.\cite{FGK}. Using the equation (3.15a) of such reference  for $L_1$   we obtain
\begin{equation}
e^{L_{1}}=1+a{\mb A}_{1}+\frac{1}{2}a^{2}(\partial_{1}{\mb A}_{1}+{\mb A}_{1}^{2})+\frac{1}{3!}a^{3}\left(\partial^{2}_{1}{\mb A}_{1}+\frac{3}{2}
\left(\frac{}{} {\mb A}_{1}\partial_{1}{\mb A}_{1}+(\partial_{1}{\mb A}_{1}){\mb A}_{1}\right)+{\mb A}_1^3\right).
\label{U1FGK}
\end{equation}
which does not agree with the expression (\ref{COMP}). This basic discrepancy propagates when combining more segments and ultimately it is the source of  the differences between our results and those obtained via the application of the methods in Ref.\cite{FGK}.

It is interesting to remark that in  the Abelian limit both (\ref{COMP}) and (\ref{U1FGK}) reduce to
\be
\label{COMPA}
U_{1}=1+a{ A}_{1}+\frac{1}{2}a^{2}(\partial_{1}{ A}_{1}+{ A}_{1}^{2})+\frac{1}{3!}a^{3}(\partial^{2}_{1}{ A}_{1}+ 3{ A}_{1}\partial_{1}{ A}_{1} + A_1^3),
\ee
which is obtained from a direct calculation using the expression (\ref{PATHI}).

\subsection{The holonomy ${\mb{h}}_{\alpha_{IJ}}$ }

Now we put the remaining pieces together in order to calculate  ${\mb{h}}_{\alpha_{IJ}}=U_3U_2U_1$. Using the notation
\be
y=s^a_J\,\partial_a
\ee
and starting from the basic structure (\ref{U1}) we obtain, mutatis mutandis,

\begin{eqnarray}
U_{2}^{(1)}&=&(s_{J}^{a}-s_{I}^{a}){\mb A}_{a}, \\
U_{2}^{(2)}&=&\frac{1}{2}[(x+y) (s_{J}^{a}-s_{I}^{a}){\mb A}_{a}+ (s_{J}^{a}-s_{I}^{a})(s_{J}^{b}-s_{I}^{b}){\mb A}_{a}{\mb A}_{b}], \\
U_{2}^{(3)}&=&\frac{1}{3!}[(x^{2}+y^{2}+xy) (s_{J}^{a}-s_{I}^{a}){\mb A}_{a}+(x+2y+\bar{y}+2\bar{x})(s_{J}^{a}-s_{I}^{a})(s_{J}^{b}-s_{I}^{b}){\mb A}_{a}\bar{{\mb A}_{b}}{}
\nonumber\\
&&{}+ (s_{J}^{a}-s_{I}^{a})(s_{J}^{b}-s_{I}^{b})(s_{J}^{c}-s_{I}^{c}){\mb A}_{a}{\mb A}_{b}{\mb A}_{c})], \\
U_{2}^{(4)}&=&\frac{1}{4!}[(x^{3}+y^{3}+x^{2}y+xy^{2})(s_{J}^{a}-s_{I}^{a}){\mb A}_{a}+(x\bar{y}+3x\bar{x}+x^{2}+2xy+2\bar{x}\bar{y}+3\bar{x}^{2}{}
\nonumber\\
&&{}+3y^{2}+\bar{y}^{2}+3y\bar{y}+5\bar{x}y)(s_{J}^{a}-s_{I}^{a})(s_{J}^{b}-s_{I}^{b}){\mb A}_{a}\bar{{\mb A}_{b}}+(x+2\bar{x}+3\bar{\bar{x}}+2\bar{y}+\bar{\bar{y}}+3y){}
\nonumber\\
&&{}\times (s_{J}^{a}-s_{I}^{a})   (s_{J}^{b}-s_{I}^{b}) (s_{J}^{c}-s_{I}^{c}){\mb A}_{a}\bar{{\mb A}_{b}}\bar{\bar{{\mb A}_{c}}}+(s_{J}^{a}-s_{I}^{a})(s_{J}^{b}-s_{I}^{b})\nonumber\\
&&(s_{J}^{c}-s_{I}^{c})(s_{J}^{d}-s_{I}^{d}){\mb A}_{a}{\mb A}_{b}{\mb A}_{c}{\mb A}_{d}],
\end{eqnarray}
for $U_2$, together with
\begin{eqnarray}
U_{3}^{(1)}&=&-s_{J}^{a}{\mb A}_{a}, \\
U_{3}^{(2)}&=&\frac{1}{2}(-y s_{J}^{a}{\mb A}_{a}+ s_{J}^{a}s_{J}^{b}{\mb A}_{a}{\mb A}_{b}), \\
U_{3}^{(3)}&=&\frac{1}{3!}(-y^{2} s_{J}^{a}{\mb A}_{a}+(2\bar{y}+y)s_{J}^{a}s_{J}^{b}{\mb A}_{a}\bar{{\mb A}_{b}}-s_{J}^{a}s_{J}^{b}s_{J}^{c}{\mb A}_{a}{\mb A}_{b}{\mb A}_{c}), \\
U_{3}^{(4)}&=&\frac{1}{4!}
\left[-y^{3}s_{J}^{a}{\mb A}_{a}+(3\bar{y}^{2}+3y\bar{y}+y^{2})s_{J}^{a}s_{J}^{b}{\mb A}_{a}\bar{{\mb A}_{b}}-(3\bar{\bar{y}}+2\bar{y}+y)s_{J}^{a}s_{J}^{b}s_{J}^{c}{\mb A}_{a}\bar{{\mb A}_{b}}\bar{\bar{{\mb A}_{c}}}\right. \nonumber \\
&& \qquad \qquad \left. +s_{J}^{a}s_{J}^{b}s_{J}^{c}s_{J}^{d}{\mb A}_{a}{\mb A}_{b}{\mb A}_{c}{\mb A}_{d}\right],
\end{eqnarray}
for $U_3$. Let us emphasize that in all the expressions above for $U_1$, $U_2$ and $U_3$, the connection is evaluated at the vertex $v$. The bars only serve to indicate the position in which the corresponding  derivative acts.

Next we write the contributions to  the holonomy in powers of the segments. According to  Eqs.(\ref{RHOL}) and (\ref{U1N})
we obtain
\be
{\mb{h}}_{\alpha_{IJ}}^{(2)}=\frac{1}{2}s_{I}^{a}s_{J}^{b}\, {\mb F}_{ab},
\label{2ORDHOL}
\ee
\begin{equation}
{\mb{h}}_{\alpha_{IJ}}^{(3)}=\frac{1}{3!}(s^{c}_I+s^{c}_J)s_{I}^{a}s^{b}_{J}{\mb D}_{c}{\mb F}_{ab},
\label{3ORDHOL}
\end{equation}

\begin{equation}
{\mb{h}}_{\alpha_{IJ}}^{(4)}=\frac{1}{4!}(s^{c}_Is^{d}_I+s_{I}^{c}s^{d}_{J}+s_{J}^{c}s^{d}_{J})s_{I}^{a}s^{b}_{J}{\mb D}_{c}{\mb D}_{d}{\mb F}_{ab}
+\frac{1}{8}s_{I}^{a}s_{J}^{b}s_{I}^{c}s_{J}^{d}{\mb F}_{ab}{\mb F}_{cd},
\label{4ORDHOL}
\end{equation}
\begin{eqnarray}
{\mb{h}}_{\alpha_{IJ}}^{(5)}&=&\frac{1}{5!}(s^{c}_Is^{d}_Is^{e}_I+s_{J}^{c}s^{d}_{J} s^{e}_J+2( s_{J}^{c}s^{d}_{I}s^{e}_I+s_{I}^{c}s^{d}_{J}s^{e}_J)-s_{I}^{c}s^{d}_{J}s^{e}_I-s_{J}^{c}s^{d}_{I}s^{e}_J)s^a_Is^b_J {\mb D}_c{\mb D}_d{\mb D}_e    \mb{F}_{ab}\nonumber\\&&
+ \frac{1}{4!}s^a_Is^b_Js^d_Is^e_J ( \mb{ F}_{ab} ( s_{I}^{c}+s^{c}_{J} ){\mb D}_c {\mb F}_{de}+(s_{I}^{c}+s^{c}_{J} )({\mb D}_c \mb{ F}_{de})\mb{ F}_{ab}).
\label{5ORDHOL}
\end{eqnarray}

Eq. (\ref{4ORDHOL}) resolves the ordering ambiguity which apparently arised in covariantizing the first term in the RHS of Eq.(\ref{MJKD4}). Nevertheless, as we subsequently show there is really no such ambiguity at this order. Let us consider the combination
\ba
s_{I}^{a}s^{b}_{J}s_{I}^{c}s^{d}_{J} \left(  {\mb D}_{c}{\mb D}_{d}-  {\mb D}_{d}{\mb D}_{c}\right){\mb F}_{ab}&=&s_{I}^{a}s^{b}_{J}s_{I}^{c}s^{d}_{J}\, \left[ {\mb D}_{c},  {\mb D}_{d}\right]{\mb F}_{ab}\nonumber \\
&=&-s_{I}^{a}s^{b}_{J}s_{I}^{c}s^{d}_{J}\, \left[ {\mb F}_{cd},{\mb F}_{ab} \right]=\left[{\mb {F}},{\mb F} \right]=0
\ea
where we have used the notation ${\mb F}=s_{I}^{a}s^{b}_{J}{\mb F}_{ab}$ together with the property
\ba
\left[ {\mb D}_{c},  {\mb D}_{d}\right]{\mb G}=-\left[{\mb F}_{cd},{\mb G} \right]
\ea
valid for any object ${\mb G}$ in the adjoint representation(Jacobi identity).

The results (\ref{2ORDHOL}), (\ref{3ORDHOL}) and  (\ref{4ORDHOL}), which we have obtained by direct calculation, 
constitute in fact the unique gauge covariant generalization of the corresponding Abelian expressions
(\ref{MJKD2}), (\ref{MJKD3}) and  (\ref{MJKD4}). This provides a strong support to our method of calculation.

\section{ The holonomy according to the method of Ref.[16]}

In this section we calculate the holonomy ${\mb{h}}_{\alpha_{IJ}}$ using the method of Ref. \cite{FGK} adapted for our case of three segments: $L_1, L_2, L_3$. From Eqs. (3.1),  (3.3), (3.4), (3.12) and (3.14), with $\lambda=1$,  of that reference it follows that
 \begin{equation}
 {\mb{h}}_{\alpha_{IJ}}=e^{L_3}e^{L_2}e^{L_1}=e^{ \Sigma+\sum_{n=2}F_{n}} =e^{H}.
\label{EQH}
\end{equation}
The basic building blocks are
\begin{equation}
L_{1}=s^{a}_{I}F(x){\mb A}_{a}(v), \quad
L_{2}=(s^{a}_{J}-s^{a}_{I})F(y-x)e^{x}{\mb A}_{a}(v), \quad
L_{3}=-s^{a}_{J}F(y){\mb A}_{a}(v).
\label{BBLOCK}
\end{equation}
where the vertex $v$ generalizes de point $(x_1, x_2)$ in the notation of
\cite{FGK}. The calculational method indicated in Eqs.(\ref{EQH}) and (\ref{BBLOCK}) is clearly not equivalent to the correct prescription (\ref{RHOL}): 
${\mb h}_{\alpha_{IJ}}=U_3\,U_2\,U_1$,
 with
the $U$'s given by Eq.(\ref{U1}) together with the corresponding extensions that take into account the change of the starting point in the corresponding path.

Let us write here those expressions arising from the method in Ref. \cite{FGK} that we will use in our calculation
\ba
\Sigma&=&L_{1}+L_{2}+L_{3}, \label{SIGMA}\\
2!F_{2}&=&[(L_{3}+L_{2}), L_{1}]+[L_{3},L_{2}], \label{F2}\\
3!F_{3}&=&-[\Sigma,F_{2}]+[(L_{3}+L_{2})^{2} L_{1}]+[L_{3}^{2}L_{2}], \label{F3}\\
4!F_{4}&=&-3![\Sigma,F_{3}]-[\Sigma^{2}F_{2}]+[(L_{3}+L_{2})^{3} L_{1}]+[L_{3}^{3}L_{2}].
\label{F4}\ea
with the notation  $[A^{2}B]=[A,[A,B]]$ and so on.

Expanding each segment in powers of the vectors $s^a$, which number is denoted by the superindex, leads to (up to third order)
\begin{equation}
L^{(1)}_{1}=s^{c}_{I}{\mb A}_{c}, \qquad
L^{(2)}_{1}=\frac{1}{2!}s^{a}_{I}s^{b}_{I}\partial_{b}{\mb A}_{a}, \qquad\label{L1}
\ee
\ba
L^{(3)}_{1}&=&\frac{1}{3!}s^{c}_{I}s^{b}_{I}s^{a}_{I}\partial_{b}\partial_{c}{\mb A}_{a},\qquad
\ea
\begin{equation}
L^{(1)}_{2}=(s^{a}_{J}-s^{a}_{I}){\mb A}_{a}, \qquad
L^{(2)}_{2}=\frac{1}{2!}(s^{b}_{I}+s^{b}_{J})(s^{a}_{J}-s^{a}_{I})\partial_{b}{\mb A}_{a}, \quad
\end{equation}
\ba
L^{(3)}_{2}&=&\frac{1}{3!}[(s^{b}_{J}s^{c}_{J}+s^{b}_{J}s^{c}_{I}+s^{b}_{I}s^{c}_{I})(s^{a}_{J}-s^{a}_{I})]\partial_{b}\partial_{c}{\mb A}_{a},
\ea
\begin{equation}
L^{(1)}_{3}=-s^{c}_{J}{\mb A}_{c}, \qquad
L^{(2)}_{3}=-\frac{1}{2!}s^{a}_{J}s^{b}_{J}\partial_{b}{\mb A}_{a},\qquad
\end{equation}
\ba
L^{(3)}_{3}&=&-\frac{1}{3!}s^{c}_{J}s^{b}_{J}s^{a}_{J}\partial_{b}\partial_{c}{\mb A}_{a}.
\label{L3}
\ea
Next we write the contributions to $H$ in powers of the segments. Using Eq.(\ref{EQH}), the first order contribution $H^{(1)}$ vanishes and the remaining contributions are
\begin{equation}
H^{(2)}=\frac{1}{2}s^{a}_{I}s^{b}_{J}{\mb F}_{ab},
\label{H2FGK}
\end{equation}
\begin{equation}
\label{H3FGK}
H^{(3)}=\frac{1}{3!}(s^{c}_{I}+s^{c}_{J})s^{a}_{I}s^{b}_{J}{\mb D}_{c}{\mb F}_{ab}
+{\mb \Sigma}^{(3)},
\end{equation}
with
\begin{equation}
{\mb \Sigma}^{(3)}=\frac{1}{12}(-s^{b}_{I}s^{a}_{I}s^{c}_{J}+s^{b}_{I}s^{a}_{J}s^{c}_{J}+s^{b}_{J}s^{a}_{I}s^{c}_{J}+s^{b}_{J}s^{a}_{J}s^{c}_{I}-s^{b}_{J}s^{a}_{I}s^{c}_{I}-s^{b}_{I}s^{a}_{J}s^{c}_{I})[\partial_{b}{\mb A}_{a},{\mb A}_{c}].
\label{NCFGK}
\end{equation}

The reader can verify that the term ${\mb \Sigma}^{(3)}$ is not 
covariant under the gauge group. This would not be the case if one had used  
the path-ordering prescription (that is to say Eq.(\ref{RHOL}) instead of Eq.(\ref{EQH})) in
the construction of the holonomies. Indeed, gauge covariance should hold to each order in the expansion in 
powers of the segments.

\section{Summary and discussion}

We have calculated the holonomy ${\mb h}_{\alpha_{IJ}}$ of the Yang-Mills connection ${\mb A}_a$ in the triangle $\alpha_{IJ}$ with vertex $v$ and sides $s^a_I$, $s^b_J$ joining at that vertex, as shown in Fig. (1). Our results, to fifth order in the segments, are given in Eqs.
(\ref{2ORDHOL}) (\ref{3ORDHOL}) (\ref{4ORDHOL}) and (\ref{5ORDHOL}) of section 4. The direct calculation shows that, to fourth order, the results are directly given by the replacement $\partial_a \rightarrow {\mb D}_a, \,\,  F_{ab}
\rightarrow {\mb F}_{ab}$ in the corresponding formulas for the Abelian case. This is so because, as explained at the end of section 4, the potential order ambiguity in the fourth order term is absent. From the fifth order on such ordering ambiguities arise, so that  is not possible to guess the correct answer from the Abelian case. Clearly then, one has to perform the full calculation in order  to obtain the correct result.

In section 5 the same  calculation was  performed using the method of Ref.\cite{FGK}.
The results for the triangle coincide to second order and start to differ from the third order on. This difference can be traced back to that arising in the calculation of the holonomy for a straight-line segment according to Eqs. (\ref{COMP}) and (\ref{U1FGK}), which the reader can easily verify. Contrary to what is expected, the calculation according to the method of Ref.\cite{FGK} produces non-covariant contributions starting at third order.

The specific results presented in Ref.\cite{FGK} correspond to the  calculation of the holonomy for a rectangle of sides  $a$ and $b$, respectively. Using the method described in section 4 we have verified that, up to fourth order, the result given in Eq. (3.19) of that reference is correct.

Nevertheless, since the calculational method of Ref. \cite{FGK} does not properly take into account the path 
ordering, it is not possible to guarantee the multiplicative composition law of holonomies. 
In fact, one might think of  obtaining the holonomy for the rectangle by composing the results of properly 
chosen triangles. Even though, as commented above, one should not expect to obtain the correct result, we have explored this possibility.
To this end let us consider a quadrilateral $ABCD$  with vertex $v$ at the point $A$, as shown in Fig. (2). The sides $AB$ and $CD$ are parallel, but $DA$ and $BC$ are not. We are interested in calculating the holonomy for the path $ABCDA$ as indicated in Fig. (2). This can be done by composing, via matrix multiplication, the holonomies corresponding to the triangles $ABC$ (spanned by the vectors ${\vec s}_I$, ${\vec s}_J$) and $ACD$ (spanned by the vectors ${\vec{ \bar{s}}}_I{}$, ${\vec{ \bar{s}}}_J{}$), each of which is calculated according to
expressions (\ref{H2FGK}) and (\ref{H3FGK}), together with their corresponding ${\mb \Sigma}$'s. In other words,

\begin{figure}
\begin{center}
\includegraphics[height=5cm]{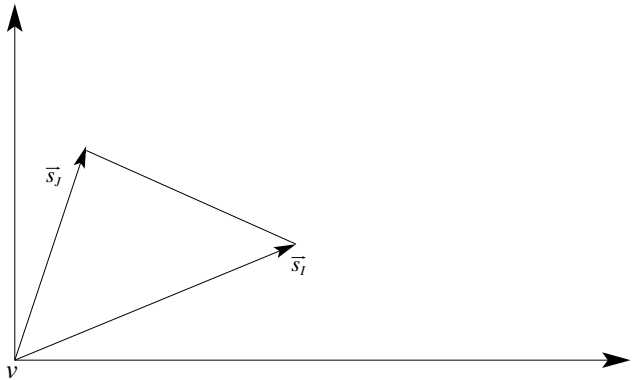}
\caption{Triangle $\alpha_{IJ}$ with vertex $v$ }
\end{center}
\end{figure}

\begin{figure}
\begin{center}
\includegraphics[height=5cm]{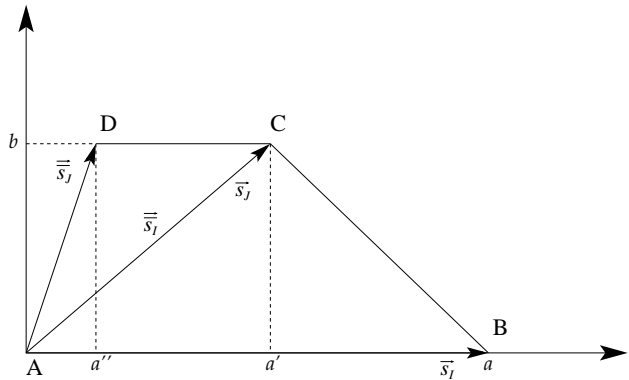}
\caption{Quadrilateral $ABCD$  with vertex $v$ at the point $A$ }
\end{center}
\end{figure}


\be
{\mb h}_{ABCDA}={\mb h}_{ACDA}{\mb h}_{ABCA}.
\label{COMPH}
\ee
We introduce the further notation in the plane of the quadrilateral
\ba
&&{\vec {AB}}={\vec s}_I= (a,0), \qquad  {\vec {AC}}={\vec s}_J= (a',b),\\
&&{\vec {AC}}={\vec{ \bar{s}}}_I{\,}= (a',b), \qquad  {\vec {AD}}={\vec{\bar{ s}}}_J{}= (a'',b).
\ea
We only pay attention to the non-covariant contributions. To third order they just add up and we obtain
\begin{eqnarray}
\label{NCFGK3}
{\mb \Sigma}_{ABCA}^{(3)}+{\mb \Sigma}_{ACDA}^{(3)}&=&\frac {(a''-a'+a)}{12}
\Big\{3a'(a''-a) [\partial_1{\mb A}_1,{\mb A}_1] \nonumber\\
&+& b (a''+a'-a) \Big([\partial_1{\mb A}_1,{\mb A}_2]+[\partial_1{\mb A}_2,{\mb A}_1]+[\partial_2{\mb A}_1,{\mb A}_1] \Big)\nonumber \\
&&  +b^2\Big[[\partial_2{\mb A}_1,{\mb A}_2]+[\partial_2{\mb A}_2,{\mb A}_1]+[\partial_1{\mb A}_2,{\mb A}_2]\Big]\Big\}.
\end{eqnarray}
We see that the above expression is not zero in general. Nevertheless, in the  symmetrical case of  a
parallelepiped, characterized by the condition
\be
a''-a'+a=0,
\ee
the non-covariant piece (\ref{NCFGK3}) vanishes. Of course, the above condition includes that of a rectangle which is  $a=a'$ y   $ a''=0 $.

Moreover, to fourth orden in the expansion, it is possible to show that following the calculational method of Ref.\,\cite{FGK},
the composition law (\ref{COMPH}) yields  gauge covariance violations even for the case of the parallelepiped.

\section*{Acknowledgements}

LFU acknowledges partial support from Fondecyt 7010967 as well as
from the project{\bf s} CONACYT 40745-F and DGAPA-IN11700.
MR acknowledges support from project (Apoyo de Tesis Doctoral)
CONICYT (Chile) and CONICYT-CONACYT 2001502170, and wants to thanks professors Luis Fernando
Urrutia and Hugo Morales-T\'ecotl for their hospitality at
ICN-UNAM and UAM. HMT acknowledges partial support from
CONICYT-CONACYT 2001502170 as well as from the project CONACYT
40745-F. The work of JA has been partially supported by Fondecyt
1010967 and CONICYT-CONACYT 2001502170.

\end{document}